\documentclass[namedreferences]{SolarPhysics}
\usepackage{graphicx}                    
\usepackage{color}                       
\usepackage{url}                         

\usepackage[pdfborder={0 0 0 },urlcolor=blue,breaklinks]{hyperref} 
\ifx \doiurl  \undefined \def \doiurl#1{\href{http://dx.doi.org/#1}{\url{#1}}}\fi
\ifx \adsurl  \undefined \def \adsurl#1{\href{http://adsabs.harvard.edu/abs/#1}{\url{#1}}}\fi

\begin{document}

\begin{article}

\begin{opening}

\title{On the Branching in the Emission Relations of Ca\,{\sc ii} lines in Solar Prominences}

%
\author{G.~\surname{Stellmacher}$^{1}$ and E.~\surname{Wiehr}$^{2}$}

%
\runningauthor{G. Stellmacher and E. Wiehr}
\runningtitle{Branching in the Ca\,{\sc ii} Emission Relations in Prominences}

%
  \institute{$^{1}$ Institute d'Astrophysique, Paris, France,
               email: \href{mailto:stell@iap.fr}{stell@iap.fr} \\
             $^{2}$  Institut f\"ur Astrophysik, G\"ottingen, Germany,
               email: \href{mailto:ewiehr@astro.physik.uni-goettingen.de} 
                          {ewiehr@astro.physik.uni-goettingen.de}\\
             }

\begin{abstract}
Spatially well resolved spectra of the emission lines Ca\,{\sc ii}\,K, 
Ca\,{\sc ii}\,8542 and H$\beta$ are analyzed in solar prominences. It is 
confirmed that the branching in the emission relations of Ca\,{\sc ii} 
versus H$\beta$ correlates with the magnitude of non-thermal (turbulent) 
broadening.
\end{abstract}

%
\keywords{Prominences, Quiescent, Ca\,{\sc ii} emission, branching}

\end{opening}

%
\section{Introduction}

High precision spectroscopy of spatially and spectrally well 
resolved emission lines in quiescent prominences indicates that 
the simple concept of an average state for prominence matter can 
only be considered as a first-order approximation. The ratio of 
the integrated line intensity ('radiance') $E(He\,D_3)/E(H\alpha)$ 
and $E(He\,D_3)/E(H\beta)$ shows a noticeable increase in fainter 
(outer) parts of prominences.

Other examples are the significant differences ('branching') 
in the relations (a) of the radiance E(Ca\,{\sc ii}\,K) vs 
E(Ca\,{\sc ii}\,8542) found by Landman and Illing (1977); 
(b) of the radiance E(Ca\,{\sc ii}\,K, H) vs $E(H\alpha, \beta$) 
found by Stellmacher (1978); and (c) of the central line intensity 
$I_0$(Ca\,{\sc ii}\,K) vs $I_0$(H$\alpha$) found by Engvold (1978). 

The branching in the Ca\,{\sc ii}\,H\&K radiance is accompanied by a 
concomitant variation of the widths of the Balmer lines relative to 
those of the Ca\,{\sc ii}\,H\&K lines (Stellmacher 1979). It has been 
conjectured whether magnetic field coupling may produce selective 
non-thermal broadening of the Ca ions. However, such an effect was 
not found in the analysis by Landman et al. (1977).

In the present work we discuss the branching in the emission relations 
of Ca\,{\sc ii}\,K on the basis of new observations of the three emission
lines Ca\,{\sc ii}\,K, H$\beta$, and Ca\,{\sc ii}\,8542 ('IR'), obtained 
at good spatial and spectral resolution.

\section{Observations}
 
A set of spectra of the three lines was obtained photographically using a 
new type of proximity focused image intensifier (PROXITRONIC, Germany; 
single-stage) in the focal plane of the Czerny-Turner spectrograph at the 
Locarno station of the G\"ottingen Observatory (cf. Wiehr et al, 1980). 
The linear dispersion amounts to 0.115, 0.145, and 0.227\,\AA/mm for 
Ca\,{\sc ii}\,K, H$\beta$ and Ca\,{\sc ii}\,IR, respectively. 

\begin{figure}[t] 
\centerline{\includegraphics[width=0.95\textwidth,clip=]{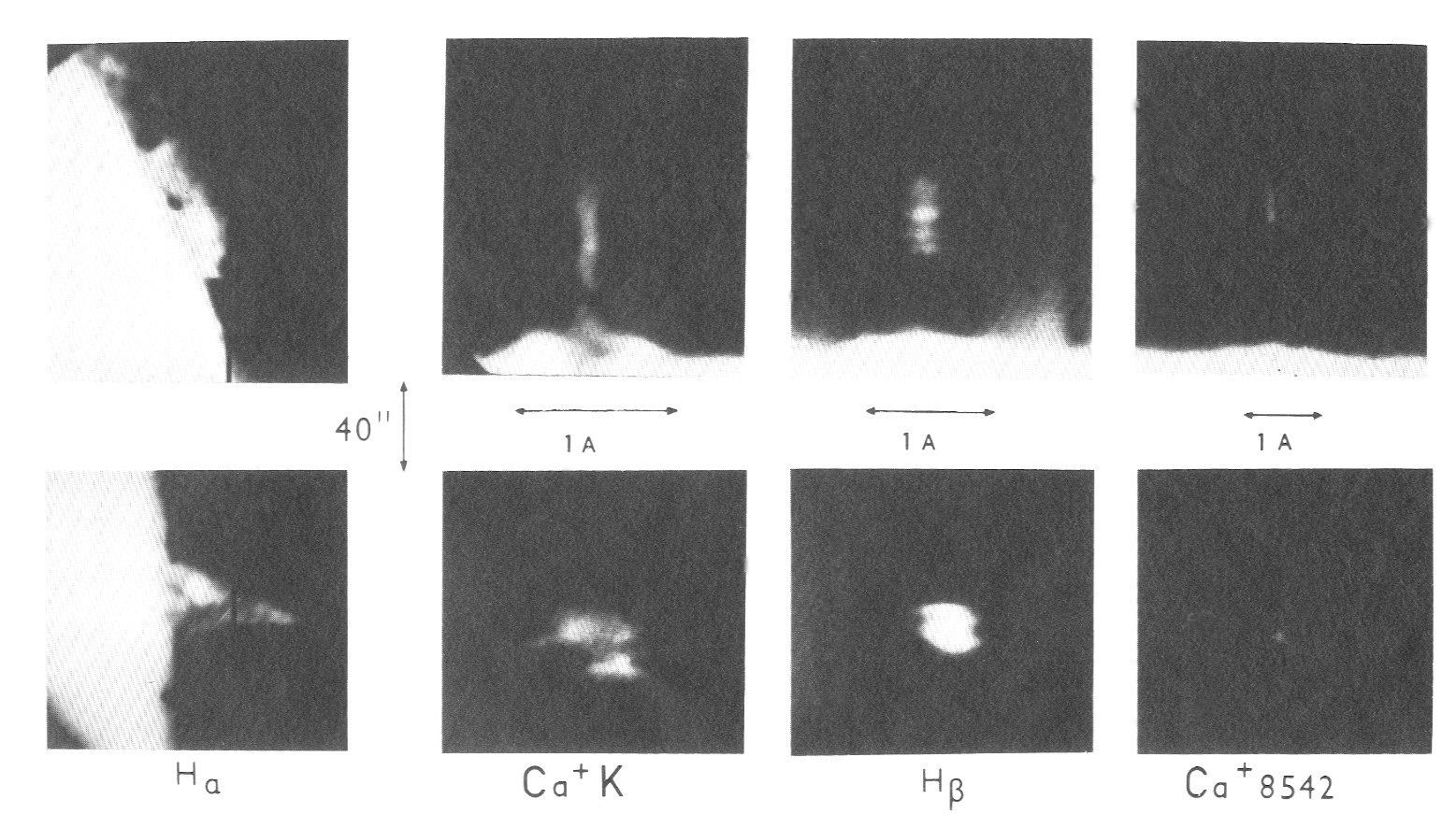}}
\caption{Two examples of observed data sets with H$\alpha$ slit-yaw 
images ({\it left panels}) and spectra of Ca\,{\sc ii}\,K, H$\beta$ and 
Ca\,{\sc ii}\,IR. Prominence E\,$40^oN$ ({\it upper}) and W\,$65^oN$
({\it lower row}).}
\end{figure}

\begin{table}[h]
\caption{Obs.\,time, heliogr.\,position and max. 
of H$\beta$ radiance [$10^4$ erg/(s\,cm$^2$\,ster)] for 
the prominences observed on July 18, 1979.}
\begin{tabular}{lcccccc} 
09:00  &   &  & east limb, $40^o$N & & & 2.65   \\
09:27  &   &  & west limb, $45^o$S & & & 0.89   \\
12:20  &   &  & west limb, $65^o$S & & & 5.01   \\ 
12:40  &   &  & east limb, $05^o$N & & & 5.28   \\
12:50  &   &  & east limb, $05^o$S & & & 9.71   \\ \\
 
\end{tabular}
\end{table}

The spectra were taken with an exposure time of $\le60$\,sec for the two 
Ca\,{\sc ii} lines, while the typical exposure time for H$\beta$ was 
$\approx12$\,sec; a total set of the three lines was obtained in 
$\approx2.5$\,min. H$\alpha$ slit-jaw images were taken for each set 
of spectra on Kodak SO-392 film. A slit width of $150\mu$ was used, 
corresponding to a spatial resolution of 1.24\,arcsec. 

For the photometric calibration of the spectra we placed a wedge in 
front of the entrance slit, illuminated by a (de-focused) quiet 
region at disk-center. The absolute prominence emission was deduced 
from disk-center spectra calibrated with the reference continua given 
by Labs and Neckel (1968). For the K-line we used the value of 0.82 
given by White and Suemoto (1968) for $\lambda=3954.2$\,\AA{} 
relative to an assumed continuum window at $\lambda=3999.9$\,\AA{}.

The micro-photometry was performed with the photometer of the Institut 
d'Astrophysique de Paris (Br\"uckner 1961). The properly scaled scattered-light 
intensity-profiles were subtracted from each scan of the prominence emission 
lines. All spectra were  observed on July 18, 1979, a day with stable seeing 
conditions. The observed prominences are listed in Table\,1. Two examples of 
a complete set of emission spectra are shown in Figure\,1.

\begin{figure}[h] 
\centerline{\includegraphics[width=0.9\textwidth,clip=]{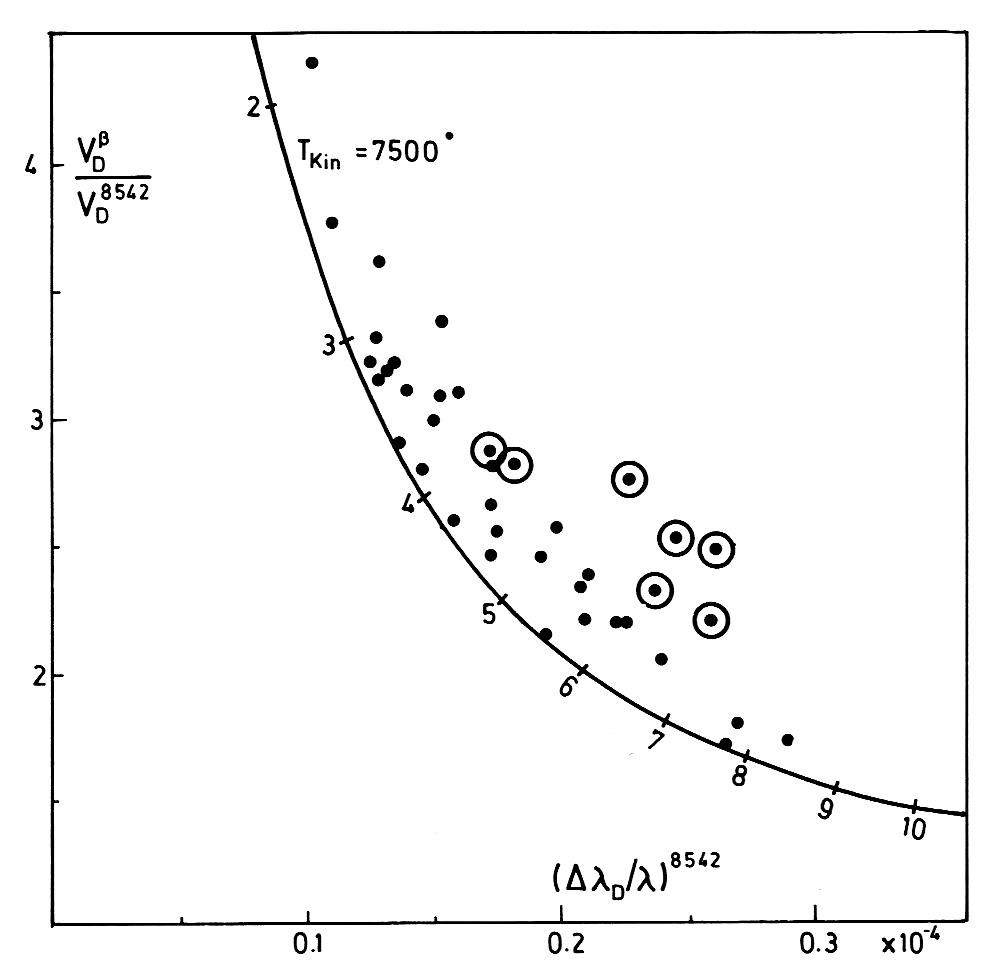}}
\caption{Observed ratio of the reduced width $v_D=\Delta\lambda/\lambda$
of H$\beta$ and Ca\,{\sc ii}\,IR vs that of Ca\,{\sc ii}\,IR. {\it Solid 
line}: calculated relation for $T_{kin}=7500$\,K and varying non-thermal
broadening given as parameter along the curve. Encircled symbols denote 
emissions with corresponding self-absorbed H$\beta$ profiles.}
\end{figure}

\section{Line Widths} 

If we suppose that for optically thin prominence emissions, the thermal 
and non-thermal line broadening constitute the only significant mechanism, 
we may deduce mean values of the kinetic temperature, $T_{kin}$, and the 
non-thermal broadening, $v_{nth}$ from the observed line profiles: 
those from atoms of small mass (e.g. H$\beta$ with $\mu=1$) depend 
mostly on $T_{kin}$, while those from atoms with large mass (e.g. Ca 
with $\mu =40$) are dominated by non-thermal broadening.

In Figure\,2 we give the ratio of the reduced Doppler widths, 
$v_D=\Delta \lambda_D/\lambda$, for  H$\beta$ and Ca\,{\sc ii}\,IR 
as a function of $v_D$(Ca\,{\sc ii}\,IR). [In the case of non Gaussian 
profiles $\Delta \lambda_D$ is defined as half width $\Delta \lambda_e$ 
at $I_o/e$.] It can be seen that the observations follow the general 
trend of a calculated curve:

$$c\cdot v_0= c\cdot\Delta\lambda_D=(2RT_{kin}/\mu + v_{nth}^2)^{-1/2}\hspace{3cm}(1)$$

\noindent
with parameters $T_{kin} = 7500$\,K and 3\,km/s$\,\le v_{nth} \le 8$\,km/s 
adapted to the lower limit of the data. These well agree with those by 
Hirayama (1978), who obtained 4500\,K$\,< T_{kin}<8500$\,K and 
3\,km/s$\,<v_{nth}<8$\,km/s. Similar values were also deduced by 
Landman et al. (1977) from their analysis of the widths of 
H$\beta$, He\,D$_3$ and Ca\,{\sc ii}\,IR.

\begin{figure}[h]
\centerline{\includegraphics[width=\textwidth,clip=]{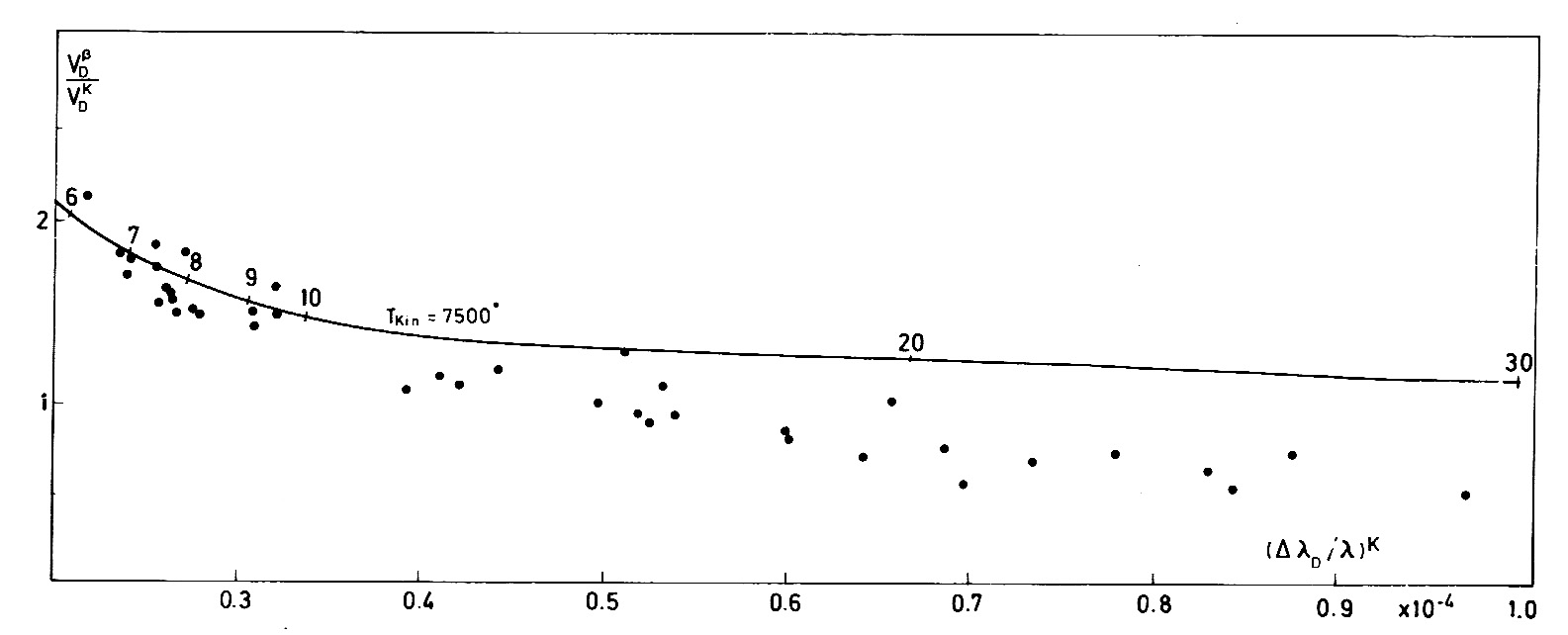}}
\caption{Observed ratio of the reduced width of H$\beta$ and Ca\,{\sc ii}\,K 
vs that of Ca\,{\sc ii}\,K. {\it Solid line}: calculated as in Fig.\,2.}
\end{figure}

A corresponding plot for the observed widths of H$\beta$ and Ca\,{\sc ii}\,K 
(Figure\,3) shows stronger deviations from the calculated curve in the sense
of additional broadening of the Ca\,{\sc ii}\,K lines. This may be due 
to self-absorption, but also to unresolved macro-shifts, visible in some 
Doppler-shifted emission ejecta (see also Engvold and Malville, 1977).

\begin{figure}
\centerline{\includegraphics[width=0.95\textwidth,clip=]{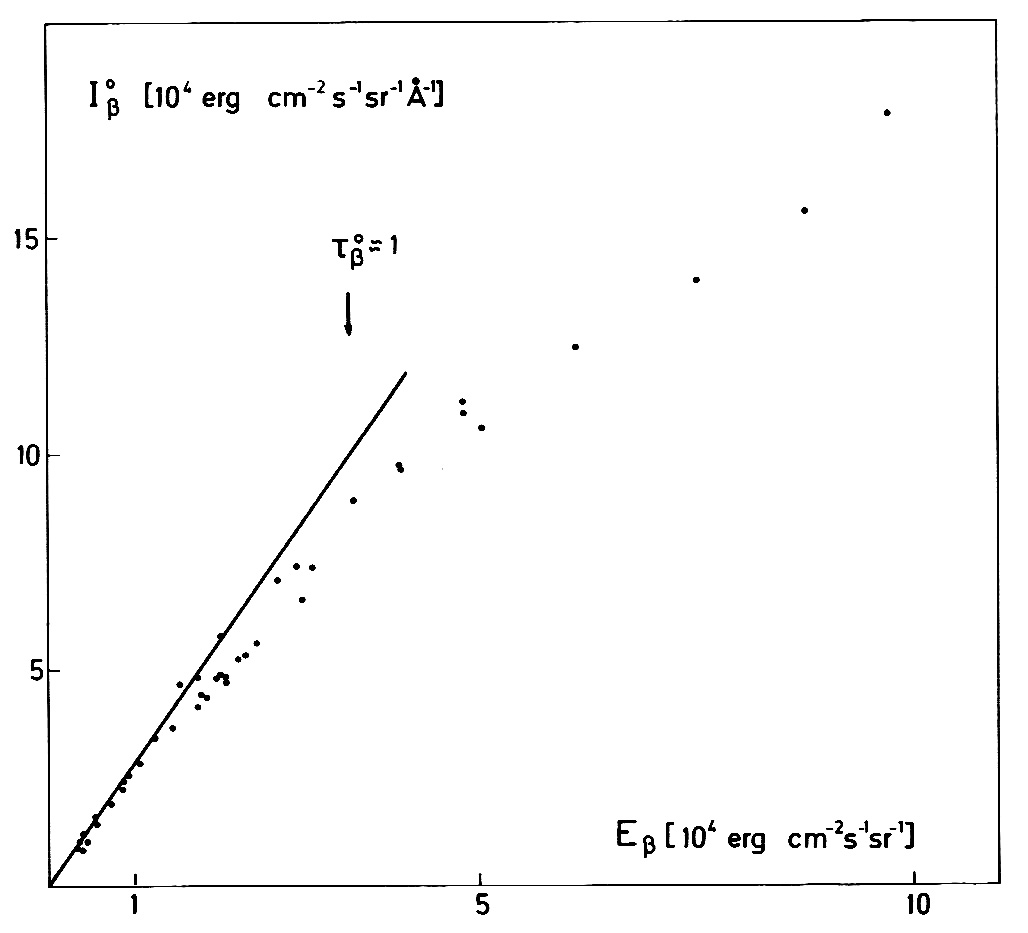}}
\caption{Observed central intensity versus integrated line emission for
H$\beta$; {\it solid line}: limiting slope corresponding to 
$\Delta\lambda_D^{min}=188$\,m\AA{}; $E(\tau_{\beta}=1)$ from Stellmacher 
(1969).}
\end{figure}

\section{Emission Relation}

The emission relations of Ca\,{\sc ii}\,H, K and H$\alpha$, H$\beta$ 
show a branching that depends on the widths of the Ca\,{\sc ii} 
lines (Stellmacher 1979). The non-thermal broadening, $v_{nth}$,
widens the profiles and reduces their central intensity, $I_0$;
the integrated intensity $E=\int{I_{\lambda} d\lambda}$ ('line radiance')
remains unchanged, except for the optically thick case.

For an optically thin line with pure Gaussian profile it is 
$E = I_0\cdot \Delta\lambda_D\cdot\sqrt{\pi}$. The relation of observed 
$I_0$ and $E$ shows for H$\beta$ (Fig.\,4) a smaller scatter than for 
Ca\,{\sc ii}\,IR (Fig.\,5). This reflects the strong dependence of the
latter on the non-thermal broadening, $v_{nth}$, which varies by more than 
a factor of three (cf., Figure\,2), whereas H$\beta$ depends mostly on the 
(much less varying) $T_{kin}$. The tight relation of $I_0^{\beta}$ and $E^{\beta}$ 
(Figure\,4) allows one to use $I_0^{\beta}$ as a measure for the total 
thickness of the prominence (i.e. its number-density).

\begin{figure}
\centerline{\includegraphics[width=0.9\textwidth,clip=]{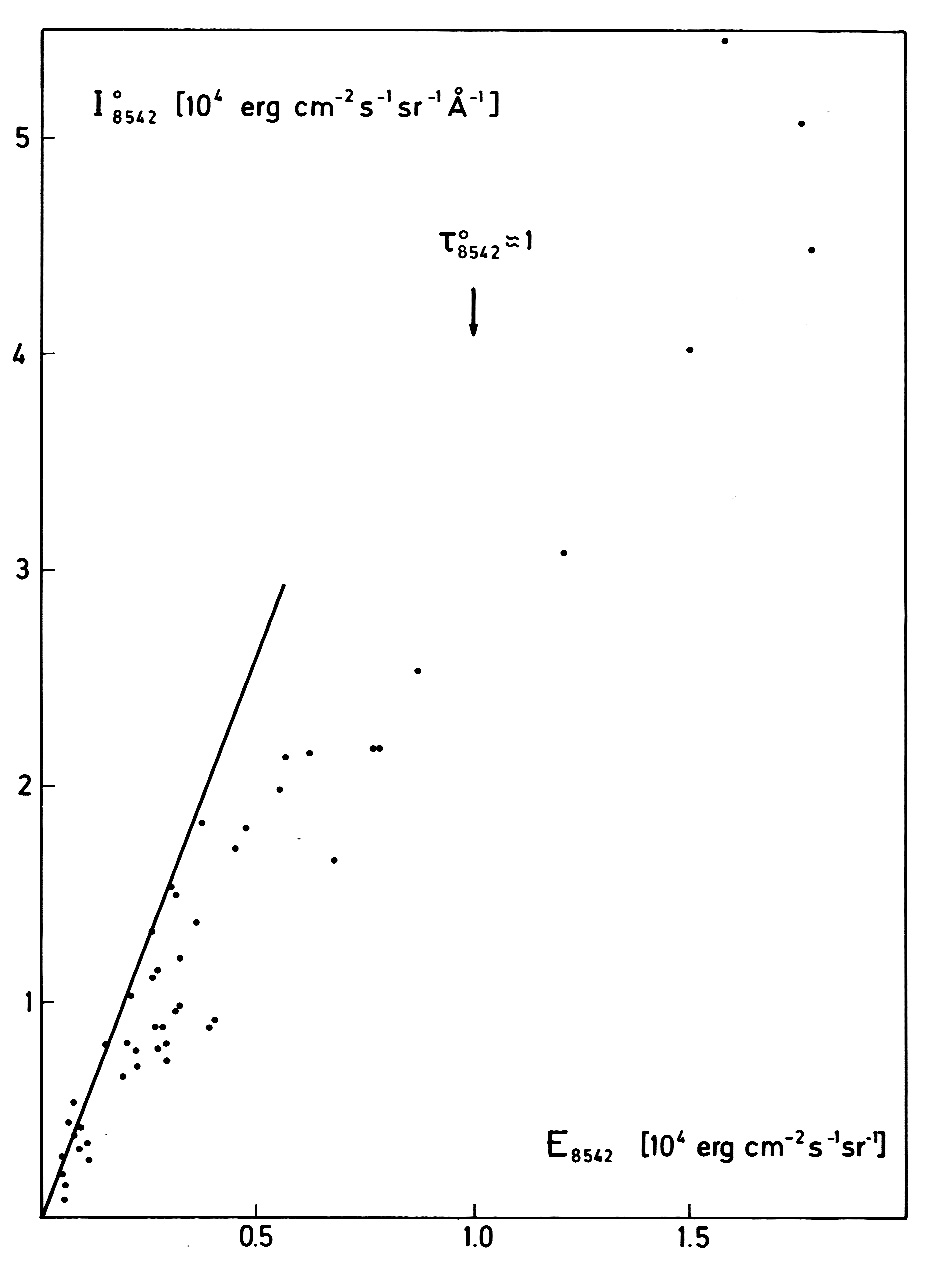}}
\caption{Observed central intensity versus integrated line emission for
Ca\,{\sc ii}\,IR; {\it solid line}: limiting slope corresponding to
$\Delta\lambda_D^{min}=105$\,m\AA{}; $E(\tau_{Ca}=1)$ from Landman (1979).}
\end{figure}

The steepest slopes in the two plots correspond to narrowest reduced 
widths $[\Delta\lambda_D^{min}/\lambda]_{H\beta}=3.87 \cdot10^{-5}$ and 
$[\Delta\lambda_D^{min}/\lambda]_{Ca IR}=1.23 \cdot10^{-5}$, respectively. 
Inserted into formula\,1 these give $T_{kin}=7570$\,K and $v_{nth}=3.6$\,km/s. 
We note that for Ca\,{\sc ii}\,IR our relation $I_0$ versus $E$ well agrees 
with the one given Landman (1979).

The bending in the observed relation $I_0$ vs E (Fig.\,4) shows 
that saturation becomes effective for the H$\beta$ emission. For an 
estimate of the H$\beta$ optical thickness, we use the H$\alpha$ 
optical thickness, $\tau_{\alpha}^0=3.6$, which Stellmacher (1969) 
gives for $E\beta\approx 1.7\cdot 10^4$ erg/(s\,cm$^2$\,ster). 
With $\tau_{\alpha}^0/\tau_{\beta}^0=(\lambda_{32}\cdot f_{32})/
(\lambda_{42}\cdot f_{42})=7.25$, we deduce a corresponding  
H$\beta$ optical thickness of $\tau_{\beta}=0.5$. Based on 
this value, we can now determine for $\tau_{\beta}= 1$ a mean 
H$\beta$ radiance $E_{\beta} = 3.5\cdot 10^4$ 
erg/(s\,cm$^2$\,ster). [Similar values are obtained on the 
basis of the analysis by Landman and Mongillo (1979).]
The widths of the seven broadest H$\beta$ lines in Figure\,4
with $E_{\beta}\ge5\cdot 10^4$ erg/(s\,cm$^2$\,ster), where 
saturation becomes important, are encircled in Figure\,2.

\begin{figure}[h]
\centerline{\includegraphics[width=1.02\textwidth,clip=]{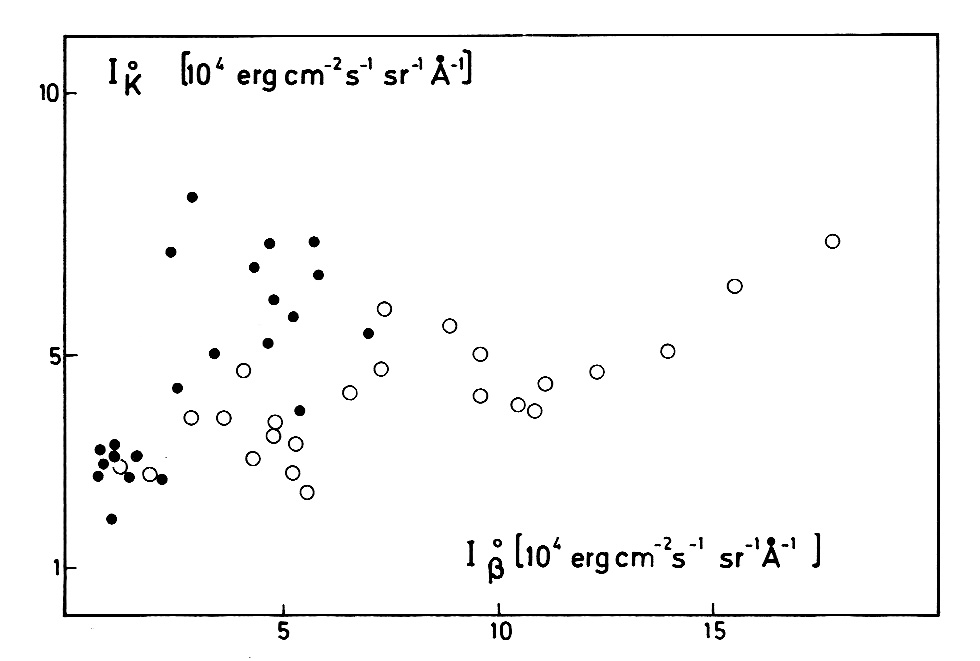}}
\caption{Observed relation of the central intensity of 
Ca\,{\sc ii}\,K and H$\beta$; open circles denote broad 
Ca\,{\sc ii}\,K profiles with $R^K<1.45$).}
\end{figure}

Figure\,6 shows the central intensity $I_0$ of the observed 
Ca\,{\sc ii}\,K line versus that of the corresponding H$\beta$ 
line, (the latter giving the total prominence thickness as 
argued above). The observed relation can be separated into two 
distinct branches defined by the ratio of their reduced Doppler 
width $v_D^{\beta}/v_D^K$ from Figure\,3. The upper branch (full 
dots in Fig.\,6) contains data with $v_D^{\beta}/v_D^K\ge 1.45$,
which are characteristic for narrow Ca\,{\sc ii}\,K profiles
($\Delta\lambda_D/\lambda < 0.3\cdot 10^{-6}$ in Fig.\,3).
The lower branch (open circles in Fig.\,6) contains data 
with $v_D^{\beta}/v_D^K\le 1.45$, which are characteristic
for broad the Ca\,{\sc ii}\,K profiles in  Figure\,3.  Hence, 
the difference in the central Ca\,{\sc ii}\,K  intensity for 
a given $I_0^{\beta}$ reflects their strong sensitivity to 
non-thermal broadening.

%
\begin{figure}
\centerline{\includegraphics[width=0.9\textwidth,clip=]{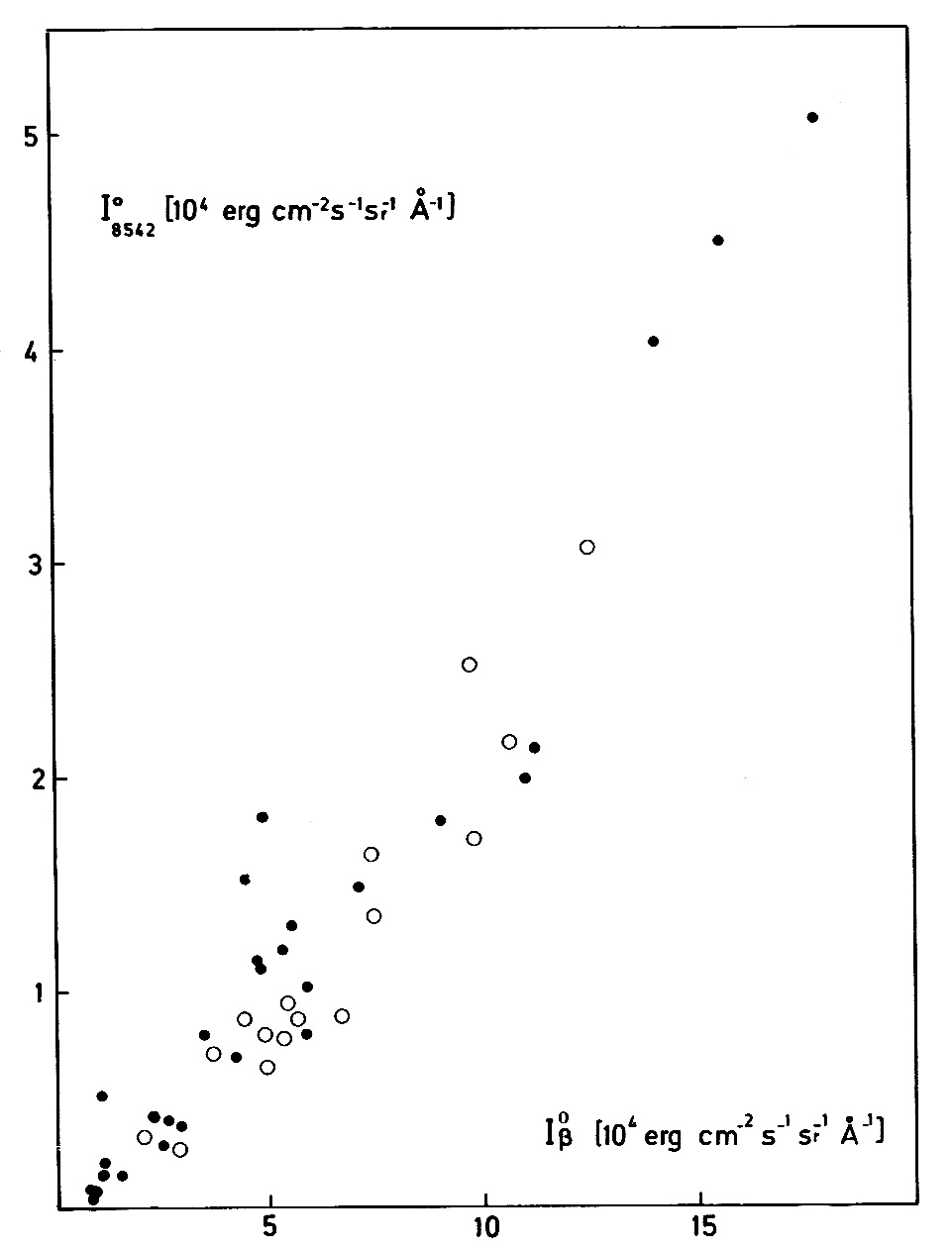}}
\caption{Observed relation of the central intensity of Ca\,{\sc ii}\,IR and
H$\beta$; open circles denote broad Ca\,{\sc ii}\,IR profiles with 
$R^K<2.5$).}
\end{figure}

A similar relation, but for the observed Ca\,{\sc ii}\,IR and 
H$\beta$ line is shown in Figure\,7. Here, the branching is 
much less pronounced. For optical thin layers with central 
intensities $I\beta\le 10\cdot 10^4$ erg/(s\,cm$^2$\,ster\,\AA) 
in Figure\,7, we find a slight indication that narrow 
Ca\,{\sc ii}\,IR profiles with $v_D^{\beta}/v_D^{IR}\ge2.5$ 
(cf., Fig.\,2) tend to follow the upper part of the relation 
(full dots in Figure\,7).

\section{Conclusions}

We confirm that the branching in the Ca\,{\sc ii} versus Balmer 
emission correlates with the non-thermal line broadening, $v_{nth}$. 
The importance of that parameter for the transfer problem of the 
Ca\,{\sc ii} lines was discussed by Heasley and Milkey (1978). 
Comparison of their calculated relation E(Ca\,{\sc ii}\,K) 
vs E(Ca\,{\sc ii}\,H) with observations by Landman and Illing 
(1977) gave small values $v_{nth}\le2$\,km/s. Stellmacher 
(1979) obtained $2\le v_{nth}\le4$\,km/s, being compatible with 
values he obtained from higher Balmer lines and He\,D$_3$.

The line-width method (through Equation (1)) may be reliable, 
if the widths of the hydrogen, helium and metal lines give 
single values for $T_{kin}$ and $v_{nth}$. However, the time
variation of the prominence structure ('threads') suggests 
that {\it large scale motions may affect a line broadening 
in addition to the thermal and non-thermal velocities}.  

Corresponding calculations have been carried out by Kawaguchi (1966) 
for the Balmer line profiles. He assumed different laws for the 
distribution of the thread velocities (independent of the position 
along the line of sight) and constant thermal and non-thermal 
(Maxwellian) broadening within each optically thin thread. He
obtained (apart from a Gaussian distribution law, which again 
results in Gaussian profiles) non-Gaussian shapes for the 
additional line broadening.  

Hence, {\it the structure of the prominence itself influences 
the line width and the total emission by convolution of the 
internal Maxwellian broadening with the (macro-) velocity 
distribution function of threads along the line of sight}. 
Side-on and edge-on observations of prominences may then 
be spectroscopically distinct; a possible example was  
discussed by Landman and Illing (1977) in context with 
the branching of their Ca\,{\sc ii}\,K versus 
Ca\,{\sc ii}\,IR emission relation. Similarly, differences 
in E(Ca\,{\sc ii}\,K)/E(H$\alpha$) ratio found by Engvold 
et al. (1978) between ordinary prominence structures 
and edge structures with 'ejecta’ may be related to such 
a modified line broadening.

%

%
%
\bibliographystyle{spr-mp-sola}
%


\end{article} 
\end{document}